\documentclass[a4paper, 12pt]{article}

\usepackage{amsmath}
\usepackage[dvips]{graphicx}

\begin{document}

\noindent\hspace{2mm}{\LARGE{\bf New $\beta$ Model of Intracluster Gas Distribution}} \\
       
\vspace{3mm}

\normalsize
\begin{flushright}
{\large Shigeru J. MIYOSHI$^1$, Masatomo YOSHIMURA$^1$,\,\,\,}\\
\end{flushright}

\vspace{-4mm}

\noindent\hspace{5.8cm}{\large and Nobuhiro TANAKA$^2$} \\

\vspace{-3mm}

\noindent\hspace{5.8cm}{\footnotesize 1. \begin{minipage}[t]{10.5cm}Department of Physics, Kyoto Sangyo University, Kyoto 603-8555, Japan\end{minipage}}\\

\vspace{-1mm}

\noindent\hspace{5.8cm}{\footnotesize 2. \begin{minipage}[t]{10.5cm}Astronomy Data Center, National Astronomical Observatory of Japan, Mitaka, Tokyo, 181-8588, Japan\end{minipage}}\\

\vspace{4mm}

\begin{center}
{\bf Abstract}\\
\end{center}

A new type of $\beta$ model of intracluster gas distribution is established, which is well fitted to the observed X-ray surface brightness profile except for outermost region. It employs the exact King model of the self-gravitating system of collisionless particles instead of the widely used approximated analytical formula for the dark matter distribution. The small difference between the modelled and the observed X-ray surface brightness profiles would disappear if the infall of matter from the intercluster space is taken into account. The number of parameters of our model is the same as for the isothermal $\beta$ model, but the agreement between our model and data points is satisfactory even when the fitting with the isothermal $\beta$ model remains a large central excess. In addition, the enhanced central mass concentration in our model is high enough to resolve the discrepancy between the X-ray and the lensing measurements of cluster mass. These facts suggest that most of the central excess might be no more than a ghost embossed by an inadequate fitting. Therefore the central excess would not necessarily be linked to the existence of cooling flow. It is also shown that the obtained values of $\beta\ ({\rm our\ model})\sim 0.5$ support the theory of non-gravitational intracluster gas heatings comparable to those reached through gravitational collapse. On the other hand, the value of $\beta\sim 0.6$ obtained for the isothermal $\beta$ model is explained as showing the scarcity of matter in the space surrounding each cluster, brobably due to the infall of matter towards it over a long time.\\

\vspace{1.5mm}

\noindent {\bf Key words:}\hspace{2mm}\begin{minipage}[t]{14.5cm}X-rays, cluster of galaxies, intracluster gas temperature, density distribution, dark matter
\end{minipage}\\

\vspace{8mm}

\begin{center}
{\large{\bf 1. Introduction}}\\
\end{center}

The standard model of mass distribution in clusters of galaxies assumes that both gas and dark matter are in hydrostatic equilibrium within the binding cluster potential. Recently many studies have compared mass measurements for clusters of galaxies using the X-ray and the lensing techniques. Miralda-Esqud\'e \& Babul (1995) reported a discrepancy by a factor of  $\sim 2$ between the X-ray and the strong-lensing mass determinations for A1689 and A2218. Wu \& Fang (1997) concluded that the mass measurements from the velocity dispersion of optical galaxies and the strong-lensing data were generally in good agreement but were typically a factor 2--3 larger than those from the X-ray data. In contrast to this, the comparison of the weak lensing and the X-ray mass measurements at cluster's outer regions has inferred good agreement between them (e.g. Squires et al. 1996; Smail et al. 1997). These facts show that the isothermal $\beta$ model widely used in the X-ray data analyses is useful preferably only in the outer region of each cluster. \par
It is therefore inebitable to modify the model of intracluster gas distribution in order to resolve the discrepancy between the X-ray and the strong-lensing mass measurements. The double $\beta$ model of Xue \& Wu (2000) is a possible one. However, as shown by themselves, the correction by the double $\beta$ model is still insufficient to resolve the discrepancy. In addition, the physical foundation of the double $\beta$ model is unclear, and the mathematical formulas to reconstruct the actual form of gas distribution are fairly tangled and unphysical. For example, the restored gas distribution is not a simple superposition of two separated components. Models of intracluster gas distribution based on the NFW model of dark matter halo having central cusp (Navarro, Frenk \& White 1996, 1997; Moore et al. 1999) or on the purely phenomenological dark matter density distribution with no central cusp (Burkert 1995) are suitable only for individually limited types of clusters. Such a state of things demands us to establish a new powerfull model of intracluster gas distribution, on the basis of firm physical foundations, which can resolve these problems all at once.\par
In this paper, we propose a new type of $\beta$ model of intracluster gas distribution in Section 2. Then we determine the parameter values of our model best fitted to observed X-ray surface brightness profiles of three clusters of galaxies and compare them with the results obtained by using the conventional isothermal $\beta$ model in Section 3. The good agreement between the X-ray and the gravitational lensing mass measurements based on our new model is also shown there. In Section 4 we discuss the new features of clusters of galaxies on the basis of our model. Brief comments are also given on the works based on other influential models. Finally, the present results are summarized in Section 5.\par
Instead of the recent {\it WMAP} results showing the existence of non-zero $\Lambda \,\,(\Omega_{\Lambda}=0.73\,\pm\,0.04)$ and the value of $H_0=72\,\pm 5\,{\rm km}\,{\rm s}^{-1}{\rm Mpc}^{-1}$ (e.g. Bennett et al. 2003; Spergel et al. 2003), we assume an Einstein-de Sitter ($\Omega_{\rm M} = 1.0, \ \Omega_{\Lambda}=0.0$) model of the universe  and adopt a Hubble constant of $H_0=50\,{\rm km}\,{\rm s}^{-1}{\rm Mpc}^{-1}$ in this paper, except for Section 4, in order to make it easy to compare the present results with many previous works which used those old cosmological parameter values. \\  

\begin{center}
\noindent{\large{\bf 2. Theoretical Foundations}}\\
\end{center}

Here we outline the theoretical basis for making up the model function of the intracluster gas distribution and check in detail the general properties of the conventional isothermal $\beta$ model. Then we establish a new type of model of intracluster gas distribution, which is expected to be more physical and more practical than the isothermal $\beta$ model.\\

\noindent{\bf 2.1 \,\,Basic relations}\\

The standard model of mass distribution in clusters of galaxies assumes that both gas and dark matter are in hydrostatic equilibrium within the cluster potential. That is,
\[\hspace{62mm}  \frac{1}{\rho_{\rm g}(r)}\frac{dP(r)}{dr}=-G\frac{M(r)}{r^2}, \hspace{55mm} (1)  \]
where $G$ is the gravitational constant, $r$ is the three-dimensional distance from the cluster center, and $\rho_{\rm g}(r)$,\,\,$ P(r)$ and $M(r)$ are the intracluster gas density, pressure, and the total mass (including dark matter) within the radius $r$, respectively. The equation of state for the intracluster gas is
\[\hspace{65mm} P(r)=\frac{\rho_{\rm g}(r)}{\mu m_{\rm H}}k_{\rm B}T_{\rm g}(r),\hspace{58mm} (2)  \]
where $m_{\rm H}$, $k_{\rm B}$, $\mu$ and $T_{\rm g}(r)$ are the hydrogen mass, the Boltzmann constant, the mean molecular weight and the temperature of intracluster gas at $r$, respectively. From Equations (1) and (2) we obtain
\[\hspace{43mm} M(r)=-\frac{k_{\rm B}T_{\rm g}(r)r}{G\mu\,m_{\rm H}}\!\left(\frac{d\ln\rho_{\rm g}(r)}{d\ln r}+\frac{d\ln T_{\rm g}(r)}{d\ln r}\right).\hspace{37mm} (3)  \]
This expression is available in good approximation even when a radial flow of intracluster gas exists, so long as the flow velocity is much less than the thermal velocity of gas particles (Miyoshi 1994). \par
On the other hand, the collisionless dark matter particles satisfy the following relation (e.g. Binny \& Tremaine 1994, hereafter BT)
\[\hspace{33mm} \frac{1}{\rho_{\rm DM}(r)}\frac{d(\rho_{\rm DM}(r)\sigma_r^2(r))}{dr}+\frac{\,2\,}{r}\left(\sigma_r^2(r)-\frac{\,1\,}{2}\sigma_t^2(r)\right)=-G\frac{M(r)}{r^2},\hspace{20mm} (4)  \]
where $\sigma_r^2(r)$ and $\sigma_t^2(r)$ are the radial and the tangential (perpendicular to the radial component) velocity dispersions, and $\rho_{\rm DM}(r)$ is the dark matter density. Equation (4) gives
\[\hspace{35mm} M(r)=-\frac{r\sigma_r^2(r)}{G}\left(\frac{d\ln\rho_{\rm DM}(r)}{d\ln r}+2\frac{d\ln\sigma_r(r)}{d\ln r}+2A(r)\right), \hspace{28mm} (5) \]
where $A(r)\equiv 1-\sigma_t^2(r)/2\sigma_r^2(r)$ indicates the degree of anisotropy in the velocity distribution of dark matter. From Equations (3) and (5) we can introduce a parameter $\beta$ defined by
\[\hspace{50mm} \beta\equiv\frac{\sigma_r^2}{\,\,\frac{\displaystyle k_{\rm B}T_{\rm g}}{\displaystyle \mu m_{\rm H}}\,\,}=\frac{\frac{\displaystyle d\ln\rho_{\rm g}}{\displaystyle d\ln r}+\frac{\displaystyle d\ln T_{\rm g}}{\displaystyle d\ln r}}{\frac{\displaystyle d\ln\rho_{\rm DM}}{\displaystyle d\ln r}+2\frac{\displaystyle d\ln\sigma_r}{\displaystyle d\ln r}+2A}. \hspace{40mm} (6) \]
When the intracluster gas is isothermal and the dark matter velocity dispersion is isotropic and $r$-independent, that is,
\[\hspace{61mm} \frac{d\ln T_{\rm g}}{d\ln r}=\frac{d\ln\sigma_r}{d\ln r}=A=0,\hspace{54mm} (7) \]
Equation (6) is reduced to
\[\hspace{62mm} \frac{d\ln\rho_{\rm g}(r)}{d\ln r}=\beta\frac{d\ln\rho_{\rm DM}(r)}{d\ln r}. \hspace{55mm} (8) \]
This is equivalent to
\[\hspace{71mm} \rho_{\rm g}(r) \propto \rho_{\rm DM}^{\beta}(r).\hspace{64mm} (9) \]

\vspace{3mm}

\noindent It should be noted here that Expression (9) is still valid in good approximation even if $A,\,\,T_{\rm g}$ and $\sigma_r$ are $r$-dependent, so long as the following inequalities hold 

\vspace{3mm}

\[\hspace{43mm} \left|\frac{d\ln T_{\rm g}}{d\ln r}\right|\!\ll\!\left|\frac{d\ln\rho_{\rm g}}{d\ln r}\right|\!, \hspace{3mm}\left|\frac{d\ln\sigma_r}{d\ln r}+A\right|\!\ll\!\left|\frac{d\ln\rho_{\rm DM}}{d\ln r}\right|\!. \hspace{33mm}(10) \]\\

\noindent{\bf 2.2 \,\,Isothermal $\beta$ model}\\

When the density distribution of dark matter is assumed to have the same form as the asymptotic density distribution of self-gravitaing isothermal gas sphere, 
\[\hspace{61mm} \rho_{\rm DM}(r) \propto [1+(r/a)^2]^{-3/2}, \hspace{51mm}(11) \]
which is fairly well fitted to the galaxy distribution in clusters of galaxies (King 1972), Expression (9) gives the isothermal $\beta$ model of intracluster gas distribution (Cavaliere \& Fusco-Femiano 1978; Jones \& Forman 1984)
\[\hspace{59mm} \rho_{\rm g}(r) = \rho_{\rm g}(0)[1+(r/a)^2]^{-\frac{3}{2}\beta},\hspace{49mm} (12) \]
where $\rho_{\rm g}(0)$ is the central gas density and $a$ is the core radius. If the intracluster gas is isothermal and the radial gas density distribution is of the form of Equation (12), the X-ray surface brightness distribution which is proportional to the integral of $\rho_{\rm g}^2(r)$ along the line of sight has the functional form of 
\[\hspace{58mm} S(R) = S(0)[1+(R/a)^2]^{\frac{1}{2}-3\beta},\hspace{48mm} (13) \]
where $S(0)$ is the central surface brightness and $R$ is the projected radial distance from the cluster center on the sky plane. Many clusters, however, are not adequately described their X-ray surface brightness profiles by this model but show some central excesses. Fabian \& Nulsen (1977) and Jones \& Forman (1984) interpreted these central excesses as being due to the cooling flow of intracluster gas. The principle of their idea is as follows. Since the X-ray emissivity is proportional to $\sqrt{T_{\rm g}}\rho_{\rm g}^2$, the lower gas temperature yields the higher X-ray emissivity, when the gas pressure $T_{\rm g}\rho_{\rm g}$ remains unchanged. The last condition is required to keep the balance between the pressure gradient and the gravitational force.\par 
Here we have to remember that the density profile of self-gravitating isothermal gas sphere with a finite value at the center has a central hump in general (see e.g. BT). The X-ray surface brightness profile simulated using this solution also has a central hump, in contrast to the isothermal $\beta$ model. It is therefore probable that the central excess in the X-ray surface brightness profile remained in the fitting with the isothermal $\beta$ model might be mostly no more than a ghost. Accordingly the central excess does not necessarily mean the existence of cooling flow. Besides, the cooling flow model cannot resolve the discrepancy between the lensing and the X-ray mass measurements mentioned in Section 1. Because, the cooling flow model requires only the enhancement of gas in the central region, and the total mass density distribution (mostly due to dark matter) remains almost unchanged. \par
These facts strongly provoke us to make up a new model function of $\rho_{\rm DM}(r)$, and therefore $\rho_{\rm g}(r)$ through Expression (9), which is expected to make clear the vagueness concerned with the central excess as well as to resolve the discrepancy between the X-ray and the lensing mass measurements. \\

\noindent{\bf 2.3 \,\,The new model we propose here}\\

Here we propose a new model function of $\rho_{\rm g}(r)$ obtained from Expression (9) by using exact King models for $\rho_{\rm DM}(r)$, which is well fitted to the observed X-ray data as shown below. \par
It is well known that the stellar distributions in globular clusters and elliptical galaxies are very well described by King models (Michie 1963; Michie \& Bodenheimer 1963; King 1966, 1981) defined by
\[\hspace{36mm} \frac{\rho (r)}{\rho_1}\!= {\rm e}^{\Psi(r) /\sigma^2}\!{\rm erf}\,(\sqrt{\Psi(r)}/\sigma) - \sqrt{\frac{4\Psi(r)}{\pi \sigma^2}}\!\left(1+ \frac{2\Psi(r)}{3\sigma^2} \right),\hspace{25mm} (14) \]
where $\rho_1$ is a normalization factor, $\sigma$ is a constant of the order of the dispersion velocity, \par
\noindent ${\rm erf}\,(x)$ is the error function defined by
\[\hspace{63mm} {\rm erf}\,(x)=\frac{2}{\sqrt{\pi}}\int_0^x {\rm e}^{-t^2}dt,\hspace{53mm} (15) \]
and $\Psi(r)$ is a solution of the following Poisson equation 
\[\hspace{16mm} \frac{1}{4\pi G \rho_1r^2}\frac{d}{d r}\!\left(r^2\frac{d\Psi(r)}{d r}\right) = - {\rm e}^{\Psi(r) /\sigma^2}{\rm erf}\,(\sqrt{\Psi(r)}/\sigma) + \sqrt{\frac{4\Psi(r)}{\pi \sigma^2}}\left(1+\frac{2\Psi(r)}{3\sigma^2} \right). \hspace{6mm} (16) \]
In terms of $\tilde{\rho} \equiv \rho /\rho_0$, $\tilde{\Psi} \equiv \Psi /\sigma^2$, and $\tilde{r} \equiv r/r_0$, where $\rho_0=\rho (0)$ and $r_0=\sqrt{9\sigma^2/4\pi G\rho_0}$ \par
\noindent (the King radius; see BT for details), Equations (14) and (16) are reduced to
\[\hspace{63mm} \tilde{\rho} (\tilde{r})= X(\tilde{\Psi}(\tilde{r})) / X(\tilde{\Psi}_0),\hspace{53mm} (17) \]
and
\[ \hspace{55mm}\frac{1}{\tilde{r}^2}\frac{d}{d\tilde{r}}\left({\tilde{r}}^2\frac{d\tilde{\Psi}(\tilde{r})}{d\tilde{r}} \right)=-9\frac{X(\tilde{\Psi}(\tilde{r}))}{X(\tilde{\Psi}_0)},\hspace{45mm} (18) \]
where $\tilde{\Psi}_0 \equiv \tilde{\Psi}(0)$, and $X(x)$ is defined by 
\[\hspace{49mm} X(x)={\rm e}^x {\rm erf}\left(\sqrt{x} \right)-\sqrt{4x/\pi}(3+2x)/3. \hspace{39mm}(19) \]
The solution of Equation (18) gives a normalized density profile $\tilde{\rho} (\tilde{r})$. The boundary condition adopted here is $d\tilde{\Psi}/d\tilde{r}=0$ at $\tilde{r}=0$. The density profiles thus obtained for finite values of $\tilde{\Psi}_0$ =\,12, 9, 6, and 3 and their projected density profiles are shown in Fig.\,4-9 of BT, each of which shows a central hump more noticeable for larger values of $\tilde{\Psi}_0$. The dispersion velocities for these solutions are nearly constant in the central region and gradually decrease with $r$ in the outer region (see Fig.\,4\,-11 of BT). Expression (10), the condition required for the validity of Equation (9), is therefore well satisfied.\par 
We can construct a new model of intracluster gas density $\rho_{\rm g}(r)$ from Equation (9) by substituting the solution $\rho (r)$ of Equation (14) for $\rho_{\rm DM}(r)$. This model is necessarily expected to have a central hump and therefore to yield some central excess in the X-ray surface brightness distribution when fitted by the isothermal $\beta$ model. The X-ray surface brightness is proportional to $S(R)$, the integral of $\rho_{\rm g}^2(r)$ along the line of sight. In Fig.~1 we plot $S(R)$ simulated for our models of $\tilde{\Psi}_0$ =\,12, 9, 6, and 3 in case of $\beta=1.0$. The expected central humps are clearly seen for large values of $\tilde{\Psi}_0$. In the limit of $\tilde{\Psi}_0 \rightarrow \infty$, $S(R)$ converges to one for the isothermal sphere, i.e. $S(R)\propto R^{-3}$ at large $R$. \\

As was pointed out by Cavaliere \& Fusco-Femiano (1978), the exact equilibrium solution can be obtained by solving the full Poisson equation containing the total (dark matter plus baryonic matter) density. In our model, however, the Poisson equation (16) or (18) includes dark matter only. That is, the gas (baryonic matter) density is not included in the right hand side of these equations. Gas particles are distributed in the gravitational potential made up by dark matter. In addition, there is no foundation on which the validity of Equation (7) is supported, and only the observational data suggest the validity of the conditions described in Expression (10). Therefore, Equation (9) should be still considered to be an approximated model of real intracluster gas distribution. Nevertheless, the agreement between our new model and the actual intracluster gas density profile is expected to be considerably improved in comparison with the isothermal $\beta$ model. Because, the fractional mass content of dark matter is overwhelming there and the gravitational field within each cluster is practically governed by dark matter. \\

\begin{center}
\noindent{\large{\bf 3. Comparison with Observations}}\\
\end{center}

We present here the results of fitting our model to the observed X-ray data of three clusters of galaxies, A\,383 (Smith et al. 2001), A\,2163 (Elbaz, Arnaud \& B\"ohringer 1995) and A\,2390 (B\"ohringer et al. 1998), compared with the results of fitting by the isothermal $\beta$ model. A\,383 and A\,2390 are classified in cD clusters, but A\,2163 is not. This cluster is classified in the Abell richness class 2. Since giant arcs are observed in the central regions of these clusters, we can also compare their projected masses, $M_{\rm proj}$, the integral of the mass density obtained from X-ray data along the line of sight within the arc radius, with their (strong) lensing masses, $M_{\rm lens}$. The strong-lensing masses of A\,383 and A\,2390 are obtained from Smith et al. (2001) and Pierre et al. (1966), respectively. For A\,2163, the strong-lensing mass determined by assuming the arcs lie at the Einstein radius is $4\times 10^{13}M_{\odot}$, but this value exceeds the weak-lensing masses obtained at some outer radii (see Fig.\,2 of Squires et al. 1997). Therefore we adopt for $M_{\rm lens}$ the value of weak-lensing mass estimated by extrapolation from outer radii.  
For the intracluster gas temperature of A\,383, Smith et al. (2001) adopted the value of $T_{\rm g}=7.1\,(\pm \,2)\,{\rm keV}$ estimated from the cooling-flow-corrected cluster $T_{\rm X}$-$L_{\rm Bol}$ relation obtained by Allen \& Fabian (1998). The use of thus obtained value of $T_{\rm g}$ is, however, not suitable for our analysis because of the large error bars accompanied with the relation. The archived {\it Chandra} data of this cluster yield the emission-weighted mean temperature to be $3.5\pm 0.2$ keV and $5.2^{+0.2}_{-0.6}$ keV for the central ($\theta_R \le 15''$, where $\theta_R$ is the angular distance from the cluster centre) and surrounding ($15'' \le \theta_R \le 120''$) regions, respectively (Furuzawa 2002). We adopt the value of $T_{\rm g}=4.0\,\,{\rm keV}$ for the present analysis. The observed redshifts ($z$), the gas temperature ($T_{\rm g}$), the projected distance of giant arc from the cluster center ($R_{\rm arc}$), and the lensing mass ($M_{\rm lens}$) of each sample cluster, we adopted for the present analysis, are given in Table~1.\\

In Figs.\,2--4 we plot the results of fitting for our model and the isothermal $\beta$ model to the observed X-ray surface brightness profiles of these clusters. The best-fit values of model parameters and the resultant $M_{\rm proj}/M_{\rm lens}$ ratios for the two models are summarized in Table 2, in which the projected mass obtained from the isothermal $\beta$ model is indicated by $M_{\rm IT, proj}$. The agreement between our model with best-fit parameter values and the observations is very satisfactory for all the sample clusters aside from their outermost regions. The differences between the model curve and data points are much less than the statistical errors for A\,383 and A\,2390. Thus our model is in general much more suitable to reproduce the X-ray surface brightness profile of galaxy cluster than the isothermal $\beta$ model, except for outermost regions. Since the disagreement between our model and observed X-ray profiles in the outermost region is considered to be due to the employment of the King model, which describes a completely isolated system whose matter density becomes zero at a finete radius, it will disappear if the infall of matter from the outer (intercluster) space is taken into account.\\

\begin{center}
\noindent{\large{\bf 4. Discussion}}\\
\end{center}

As shown above, our model is well fitted to the X-ray surface brightness profile except for outermost region. The fit is satisfactory even when the fit by the isothermal $\beta$ model remains a large central excess (A\,383 and A\,2390). It should be noted here that our new model is well fitted to observed results for all of the sample clusters of galaxies (including both cD and non-cD types). The agreement between $M_{\rm proj}$ and $M_{\rm lens}$ also looks like complete. The central mass concentration in our model is high enough to resolve the discrepancy between X-ray and lensing mass measurements. The ratio of the projected mass obtained using our model to the lensing mass is ranging from 0.92 to 1.06 for the three sample clusters, while the ratio obtained using the isothermal model ranges from 0.46 to 0.94 (see Table~2). This shows the predominance of our model over the isothermal $\beta$ model. The agreement between $M_{\rm proj}$ and $M_{\rm lens}$ would be improved if we take into account the infall of matter from the intercluster space. Since our model is based on firm physical foundations, these facts suggest that most of the `central excess' of the X-ray surface brightness obtained so far would probably be no more than a ghost embossed by the use of inadequate model function. Therefore the central excess should not be directly linked to the cooling flow. 

Table 2 gives $\beta = 0.53\pm 0.08$ and $0.63\pm 0.02$ for our model and the isothermal $\beta$ model, respectively. If the velocity dispersion of intracluster dark matter is isotropic and if the intracluster gas has experienced no non-gravitational heating, $\beta$ should not be less than unity because of the intracluster gas cooling by X-ray emissions (see Equation (6)). Therefore, the values of $\beta\,({\rm our\,model})\sim 0.5$ obtained here support the theory of non-gravitational intracluster gas heatings comparable to those reached through gravitational collapse (Kaiser 1966; Evrard \& Henry 1991; Ponman, Cannon \& Navarro 1999). The large value of $\beta\,({\rm our\,model})$ for A\,2163 is supposed to show the existence of some collective motions of intracluster gas. Indeed this cluster is not relaxed but just in the process of formation (Squires et al. 1997).\par
So far we estimated the sizes and masses of sample clusters by assuming an Einstein-de Sitter ($\Omega_{\rm M} = 1.0, \,\Omega_{\Lambda}=0.0$) model of the universe for the convenience of the comparison of our results with other previous works. Here we consider the more realistic $\Omega_{\rm M}=0.3$ and $\Omega_{\Lambda}=0.7$ cosmology too. Because, the $r$-dependence of the infalling matter density is sensitive to the value of $\Omega_{\rm M}$. Gunn \& Gott III (1972) have calculated the infall of matter into the Coma cluster using the top hat model. Their calculations show that the infalling matter density varies as $\rho\propto r^{-1.67}$ for $\Omega_{\rm M}=2.0$, $\propto r^{-1.55}$ for $\Omega_{\rm M}=0.5$, and $\propto r^{-1.5}$ for $\Omega_{\rm M}=0.2$. Accordingly we adopt $\rho\propto r^{-1.52}$ for $\Omega_{\rm M}=0.3$, which gives the integral of the square of infalling matter density along the line of sight varying as $R^{1-2\times 1.52}=R^{-2.04}$. It is noted here that this $R$-dependence will be valid for both of dark and baryonic matter particles, because the influence of gas pressure is negligible at the outermost region. In order to see the actual $R$-dependence at large $R$ (outer region) it is convenient to use the results of fit using the ispthermal $\beta$ model, which gives $S(R)\propto R^{1-6\beta
}$ there. The exponent --2.04 of $R$ corresponds to the value of $\beta=0.51$ of the isothermal $\beta$ model. This value is less than the observed mean value of $\beta$\,(isothermal) obtained here for the sample clusters and the mean value for 33 clusters (by the single $\beta$ fit), $<\beta>=0.61\pm 0.09$ obtained by Xue \& Wu (2000). These facts show that the matter density at cluster's outer region decreases with $r$ more rapidly than estimated by the top hat model. This implies that the matter density in the space surrounding each cluster is fairly lower than the theoretically estimated intercluster mean matter density, probably due to the infall of matter towards each cluster over a long time. The recent discovery of a completely empty void of $\sim 140$\,Mpc radius (Rudnick et al. 2007) seems to support this point of view. \par
As was mentioned in Subsection 2.1, Expression (9) is available even if $A, \,T_{\rm g} \,{\rm and} \,\sigma_r$ are $r$-dependent so long as Expression (10) holds. In such a case, however, $\beta$ would not have a constant value throughout a cluster, because $T_{\rm g}$ and $\sigma_r$ are expected to have different $r$-dependences in general. Now, from the {\it Chandra} or XMM-Newton observations, all of the sample clusters are found to have the gas temperature descent in the central region. The rates of temperature descent are respectively $\sim 20\%$ for A\,383 (Furuzawa 2002), $\sim 30\%$ for A\,2163 (Markevitch \& Vikhlinin 2001) and $\sim 50\%$ for A\,2390 (Allen, Ettori \& Fabian 2001). For each of these sample clusters, however, a very good fit of our model to the observed X-ray profile is obtained for $r$-independent values of $T_{\rm g}$ and $\beta$. Probably, the small change in the X-ray surface brightness caused by the gas temperature descent and the contributions of cD galaxy (when it exists) in the central region might be well compensated by a little but suitable adjustment for the best-fit values of $\tilde{\Psi}_0$ and $\beta$. The parameter values compiled in Table 2 should therefore be understood as averaged over the whole system of each cluster. \par
According to an attempt to derive the intracluster gas distribution embedded in a dark matter halo with an NFW density profile (Makino, Sasaki \& Suto 1998), the obtained gas density profile is well approximated by the conventional $\beta$ model (see also Suto, Sasaki \& Makino 1998; Wu \& Chiueh 2001; Wu \& Xue 2000). In contrast to our model, however, some extra component is needed in case of the NFW dark halo to reproduce the central excess in the X-ray surface brightness distribution. The basic difference between our model and the models based on the NFW model is focused on whether or not the central cusp exists for the dark matter distribution. This problem is still unsettled, and more detailed theoretical and observational studies are needed to bring this problem to an end.\par
Aguirre, Schaye \& Quataert (2001) and Sanders (2003) have tried to reproduce the observed distribution of intracluster gas on the basis of the modified Newtonian dynamics (MOND, Milgrom 1983a,\,b,\,c) and showed that some additional dark matter component is still needed to get a good agreement between the modelled and observed results, especially in the central region of cluster. That is, the problem of the enhanced central mass concentration in clusters of galaxies is in principle separated from that concerning the validity of MOND.\\

\begin{center}
\noindent{\large{\bf 5. Conclusions}}\\
\end{center}

A new model of intracluster gas distribution is established here. It is a new type of $\beta$ model well fitted to the X-ray surface brightness profile except for outermost region of each cluster. The little gap between the modelled and observed X-ray surface profiles would disappear if the infall of matter from the intercluster space is taken into account. The fitting is satisfactory even when the X-ray surface brightness profile fitted to the isothermal $\beta$ model remains a large central excess. The enhanced central mass concentration in our model is almost high enough to resolve the discrepancy between the X-ray and the gravitational lensing mass measurements. A little sink of the modelled X-ray surface brightness at the outermost region in our model would disappear if the infall of matter from the intercluster space is taken into account. These facts suggests that the central excess might be no more than a ghost embossed by an inadequate fitting and that the central excess should not be directly connected to the existence of cooling flow. \par
The values of $\beta\,({\rm our\ model})\sim 0.5$ obtained here support the theory of non-gravitational intracluster gas heatings comparable to those reached through gravitational collapse. On the other hand, the value of $\beta\sim 0.6$ obtained for the isothermal $\beta$ model is explained as showing the scarcity of matter in the space surrounding each cluster, probably due to the infall of matter towards it over a long time.\\

\begin{center}
\noindent{\large{\bf Acknowledgments}}\\
\end{center}

This work was supported in part by the Grant-in-Aid for Scientific Research (C) of Japan Society for the Promotion of Science (No.\,14540230). It was also supported in part by the Institute for Comprehensive Research, Kyoto Sangyo University.\\

\begin{center}
\noindent{\large{\bf References}}\\
\end{center}

\footnotesize

\noindent Aguirre, A., Schaye, J. \& Quataert, E. 2001, ApJ, 561, 550.\par
\noindent Allen, S. W., Ettori, S. \& Fabian, A. C. 2001, MNRAS, 324, 877.\par
\noindent Allen, S. W. \& Fabian, A. C. 1998, MNRAS, 297, L57.\par
\noindent Bahcall, N. A. \& Lubin, L. M. 1994, ApJ, 426, 513.\par
\noindent Bennett, C. L., Halpern, M., Hinshaw, G., Jarosik, N., Kogut, A., Limon, M., Meyer, S. S., Page, L., Spergel, \par
\noindent\hspace{10mm}D. N., Tucker, G. S., Wollack, E., Wright, E. L., Barns, C., Greason, M. R., Hill, R. S., Komatsu, E.,\par
\noindent\hspace{10mm}Nolta, M. R., Odegard, N., Peiris, H. V., Verde, L. \& Weiland, J. L. 2003, ApJS, 148, 1.\par 
\noindent Binny, J. \& Tremaine, S. 1987, Galactic Dynamics, Princeton Univ. Press, Princeton (BT).\par
\noindent B\"ohringer, H., Tanaka, Y., Mushotzky, R. F., Ikebe, Y. \& Hattori, M. 1998, A\&A, 334, 789.\par
\noindent Burkert, A. 1995, ApJ, L25.\par
\noindent Cavaliere, A. \& Fusco-Femiano, R. 1978, A\&A, 70, 677.\par
\noindent Elbaz, D., Arnaud, M. \& B\"ohringer, H. 1995, A\&A, 293, 337.\par
\noindent Evrard, A. E. \& Henry, J. P. 1995, ApJ, 383, 95.\par
\noindent Fabian, A. C. \& Nulsen, P. J. E. 1977, MNRAS, 180, 479.\par
\noindent Furuzawa, A. 2002, private communications.\par
\noindent Gunn, J. E. \& Gott III, J. R. 1972, ApJ, 176, 1.\par
\noindent Jones, C. \& Forman, W. 1984, ApJ, 276, 38.\par
\noindent Kaisar, N. 1991, ApJ, 383, 104.\par
\noindent King, I. R. 1966, AJ, 71, 64.\par
\noindent King, I. R. 1972, ApJ, 174, L123.\par
\noindent King, I. R. 1981, QJRAS, 22, 227.\par
\noindent Makino, N., Sasaki, S. \& Suto, Y. 1998, ApJ, 497, 555.\par
\noindent Markevitch, M. \& Vikhlinin, A. 2001, ApJ, 563, 95.\par
\noindent Michie, R. H. 1963, MNRAS, 125, 127.\par
\noindent Michie, R. H. \& Bodenheimer, P. H. 1963, MNRAS, 126, 269.\par  
\noindent Milgrom, M. 1983a, ApJ, 270, 365.\par
\noindent Milgrom, M. 1983b, ApJ, 270, 371.\par
\noindent Milgrom, M. 1983c, ApJ, 270, 384.\par
\noindent Miralda-Esqud\'e, J. \& Babul, A. 1995, ApJ, 449, 18.\par 
\noindent Miyoshi, S. J. 1994, in Clusters of Galaxies, eds. Durret, F. A., Mazure A. \& Tr\^an Thanh V\^an, J., Editions\par \noindent\hspace{10mm}Frontieres, France, p. 187.\par 
\noindent Moore, B., Quinn, T., Governato, F., Stadel, J. \& Lake, G. 1999, MNRAS, 1147.\par
\noindent Navarro, J. F., Frenk, C. S. \& White, S. D. M. 1996, ApJ, 462, 563.\par
\noindent Navarro, J. F., Frenk, C. S. \& White, S. D. M. 1997, ApJ, 490, 493.\par
\noindent Pierre, M., Le Borgne, J. F., Soucail, G. \& Kneib, J. P. 1996, A \& A, 311, 413.\par
\noindent Ponman, T. J., Cannon, D. B. \& Navarro, J. F. 1999, Nature, 397, 135.\par
\noindent Rudnick, L., Brown, S. \& Williams, L. R. 2007, ApJ, 671, 
40.\par
\noindent Sanders, R. H. 2003, MNRAS, 342, 901.\par
\noindent Smail, I., Ellis, R. S., Dressler, A., Couch, W. J., Oemler, A., Sharples, R. M. \& Butcher, H. 1997, ApJ, 479,\par
\noindent\hspace{10mm}70.\par
\noindent Smith, G. P., Kneib, J.-P., Ebeling, H., Czoske, O. \& Smail, I. 2001, ApJ, 552, 493.\par 
\noindent Spergel, D. N., Verde, L., Peiris, H. V., Komatsu, E., Nolta, M. R., Bennett, C. L., Halpen, M., Hinshaw, G.,\par
\noindent\hspace{10mm}Jarosik, N., Kogut, A., Limon, M., Meyer, S. S., Page, L., Tucker, G. S., Weiland, J. L., Wollack, E.\par
\noindent\hspace{10mm}\& Wright, E. L. 2002, ApJS, 148, 97.\par
\noindent Squires, G., Kaiser, N., Babul, A., Fahlman, G., Woods, D., Neumann, D. M. \& B\"ohringer, H. 1996, ApJ, 461, \par
\noindent\hspace{10mm}572.\par
\noindent Squires, G., Neumann, D. M., Kaiser, N., Arnaud, M., Babul, A., B\"ohringer, H., Fahlman, G. \& Woods, D.\par
\noindent\hspace{10mm}1997, ApJ, 482, 648.\par
\noindent Suto, Y., Sasaki, S. \& Makino, N. 1998, ApJ, 509, 544.\par
\noindent Wu, X.-P. \& Chiueh, T. 2001, ApJ, 547, 82.\par
\noindent Wu, X.-P. \& Fang L. Z. 1997, ApJ, 483, 62.\par
\noindent Wu, X.-P. \& Xue Y.-J. 2000, ApJ, 542, 578.\par
\noindent Xue, Y.-J. \& Wu X.-P. 2000, MNRAS, 318, 715.\par

\newpage

\noindent{\bf Table 1.}\,\,\,Redshifts, gas temperatures, arc radii, and \par  
\noindent lensing masses of sample clusters of galaxies.\\
\begin{tabular}{lccrc}
\hline
Cluster & $z$ & $T_{\rm g}$ & $R_{\rm arc}$ & \,\,\,\,$M_{\rm lens}$ \\
        &  & (keV) & (kpc) & \,\,\,\,$(10^{13}M_{\odot})$            \\                
\hline 
A\,383  & \hspace{2mm} 0.188\hspace{2mm}  & 4.0\hspace{3.5mm} & 65\,\, & \hspace{0.5mm}4.0\,$^{+\,1.1}_{-\,1.7}$ \\
A\,2163 & \hspace{2mm} 0.201\hspace{2mm}  & $14.6^{+0.9}_{-0.8}$ & 66\,\, & \hspace{0.5mm} 2.4\,$\pm \,0.8$ \\
A\,2390 & \hspace{2mm} 0.231\hspace{2mm}  & \hspace{2mm}$11.1\!\pm\!1.0$ & 177\,\, & \hspace{0.8mm} 16\hspace{3mm}$\pm \,2$ \hspace{3.8mm} \\
\hline
\end{tabular}\\

\vspace{4cm}

\footnotesize

\noindent\hspace{-1.3cm}
\begin{minipage}[t]{17.5cm}{\bf Table 2.} Best-fit parameter values for our model and isothermal $\beta$ model, and the resultant projected mass and projected-to-lensing mass ratio within the arc radius of each cluster.\\
 \begin{tabular}{@{}lccccccccc}
 \hline
 
 \footnotesize
 Cluster & $\tilde{\Psi}_0$ & $\beta$ & $\beta$ & $r_0$ & $a$ & $M_{\rm proj}$ & $M_{\rm IT, proj}$ & 
 \begin{minipage}[b]{10mm}
 \[ \frac{M_{\rm proj}}{M_{\rm lens}} \]
 \end{minipage} &
 \begin{minipage}[b]{15mm}
 \[ \frac{M_{\rm IT, proj}}{M_{\rm lens}} \]
 \end{minipage}\\
   &  & (our model) & (isothermal) & (kpc) & (kpc) & ($10^{13}M_{\odot}$) & ($10^{13}M_{\odot}$) &  &   \\
\hline
\vspace{0.5mm}
 A\,383 & $8.0_{-0.3}^{+0.4}$ & $0.54\pm 0.02$ & $0.65\pm 0.03$ & $36\pm 3$ & $93\pm 12$ & $3.8\pm 0.2$ & $1.7\pm 0.2$ & $0.95\pm 0.31$ & $0.46\pm 0.15$ \\
 
\vspace{0.5mm}
 
A\,2163 & $7.4_{-0.2}^{+0.4}$ & $0.60\pm 0.01$ & $0.62\pm 0.02$ & $310_{-19}^{+29}$\hspace{2mm} & $305\pm 19$\hspace{2mm} & $2.2\pm 0.2$ & $2.2\pm 0.2$ & $0.92\pm 0.34$ & $0.92\pm 0.34$ \\


A\,2390 & \hspace{2.4mm}$8.2\pm 0.2$ & $0.45\pm 0.02$ & $0.62\pm 0.01$ & $39_{-2}^{+10}$ & $183\pm 7$\hspace{4mm} & 17\hspace{3.1mm}$\pm \,2$\hspace{4.1mm} & 15\hspace{3.1mm}$\pm \,1$\hspace{4mm} & $1.06\pm 0.17$ & $0.94\pm 0.14$  \\
\hline 
\end{tabular}
\end{minipage}

\newpage

\normalsize

\begin{figure}[htbp]
 \begin{center}
  \includegraphics[width=180mm]{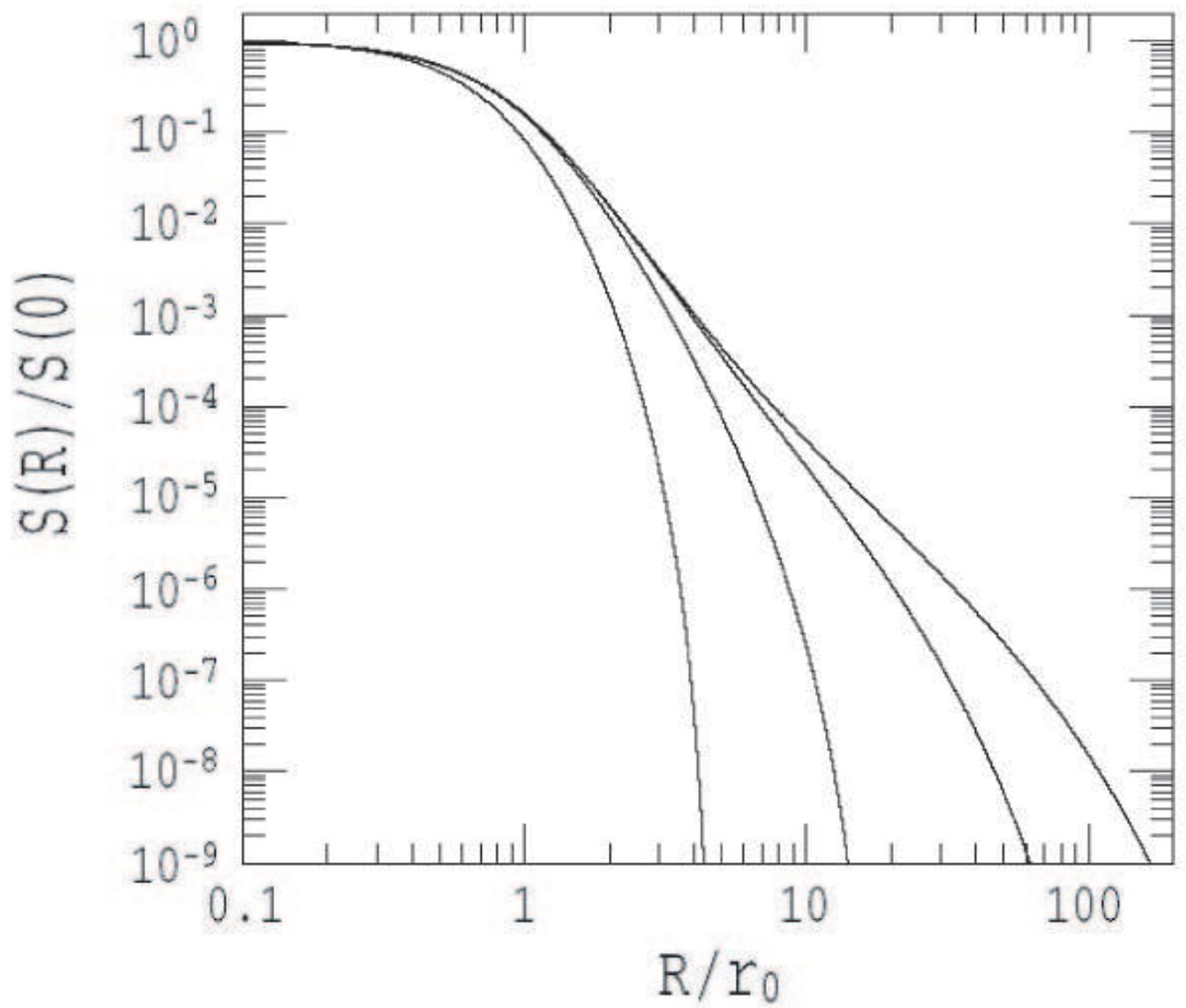}
 \end{center}
 \caption{Simulated X-ray surface brightness plofiles for our model in case of $\beta=1.0$: from top to bottom
 the central potentials of these models satisfy $\tilde{\Psi}_0$ =\,12, 9, 6, and 3.}
 \label{fig:one}
\end{figure}

\begin{figure}[htbp]
 \begin{center}
  \includegraphics[width=180mm]{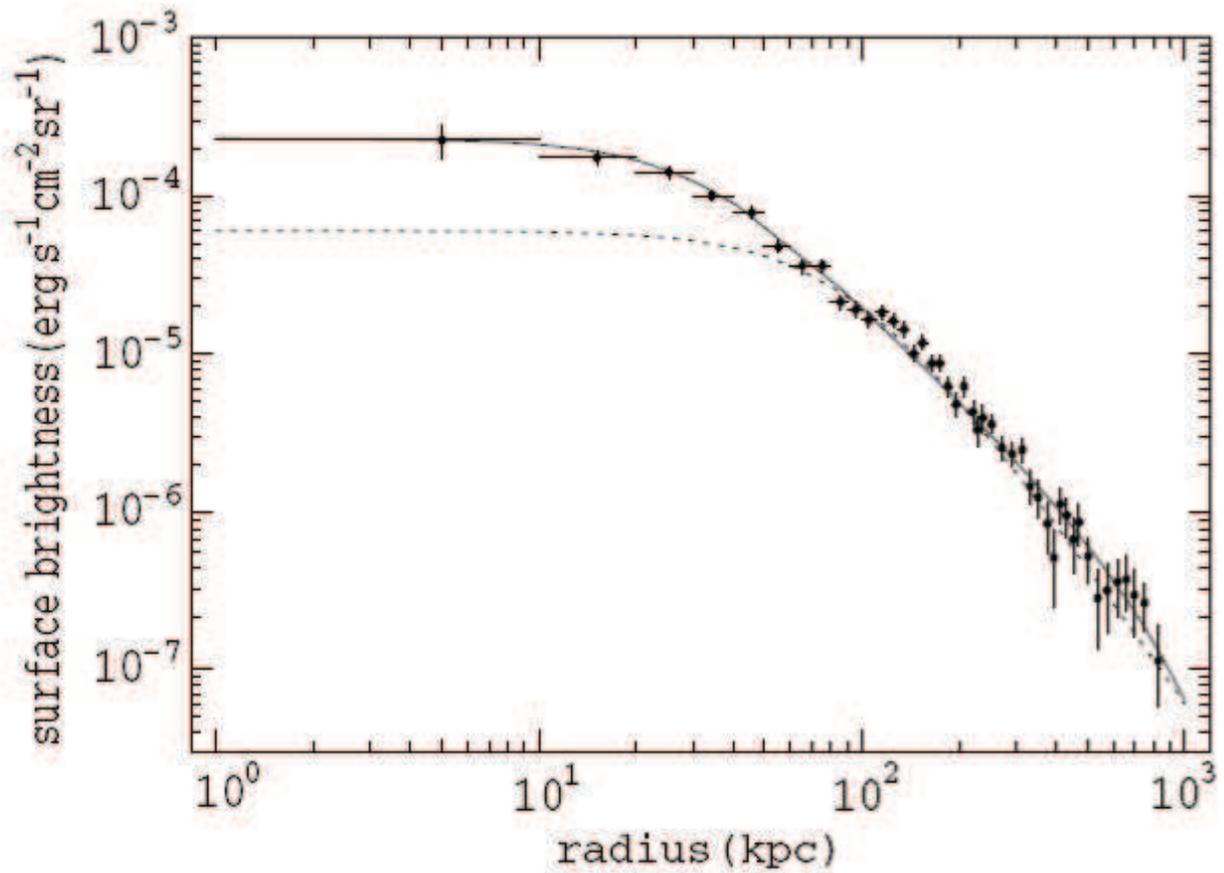}
 \end{center}
 \caption{X-ray surface brightness plofile of A\,383 (Smith et al. 2001) fitted by our
 model (solid line) and the isothermal $\beta$ model (dushed line).}
 \label{fig:two}
\end{figure}

\begin{figure}[htbp]
 \begin{center}
  \includegraphics[width=120mm,angle=90]{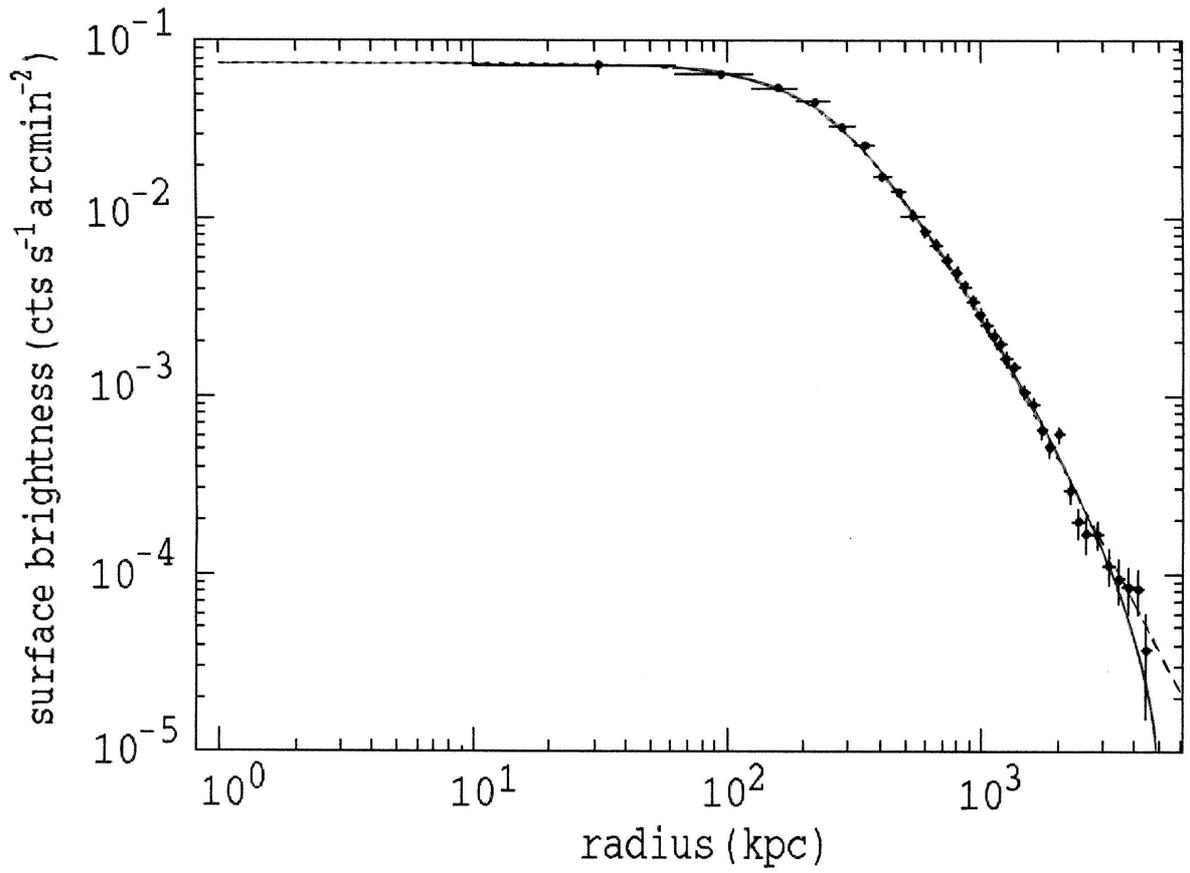}
 \end{center}
 \caption{X-ray surface brightness plofile of A\,2163 (Elbaz, Arnaud \& B\"ohringer 1995) fitted by
 our model (solid line) and the isothermal $\beta$ model (dushed line).}
 \label{fig:three}
\end{figure}

\begin{figure}[htbp]
 \begin{center}
  \includegraphics[width=180mm]{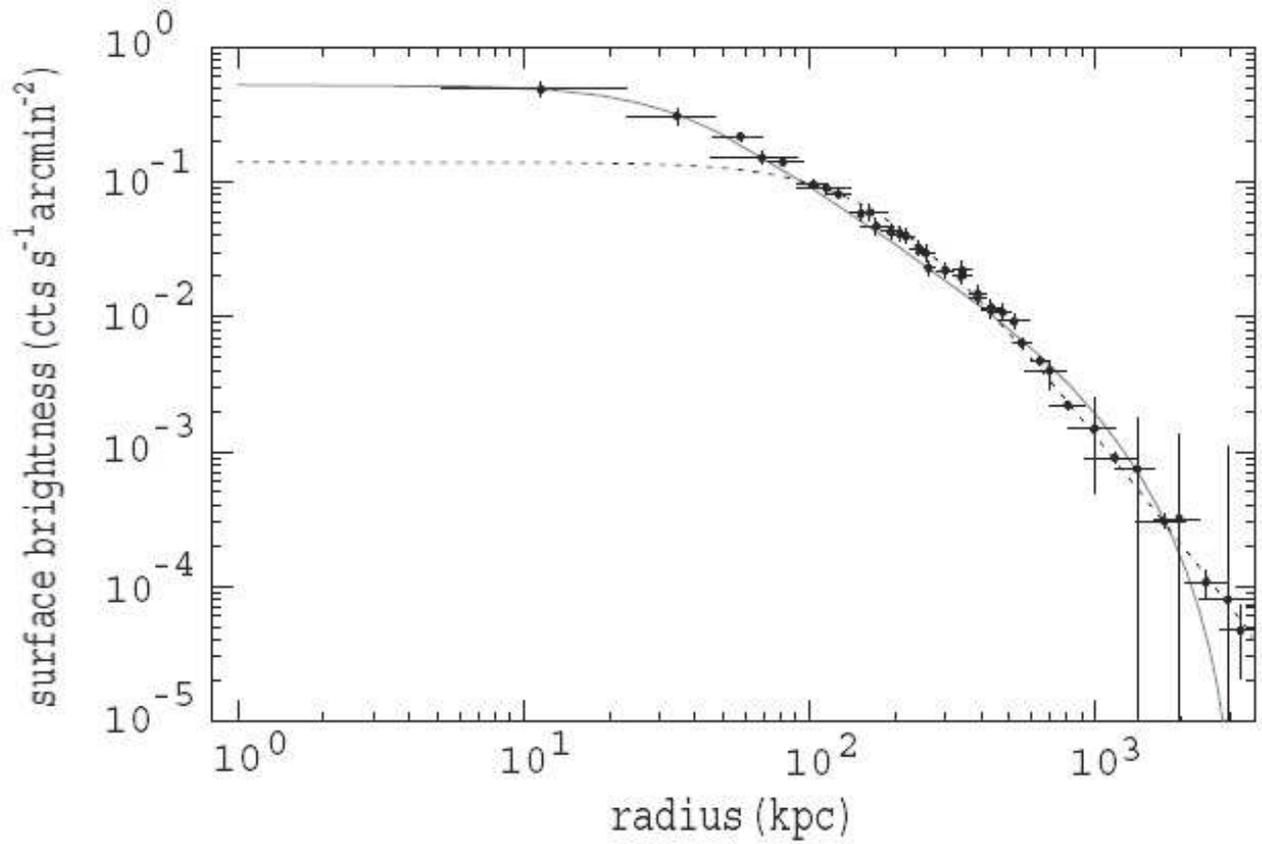}
 \end{center}
 \caption{X-ray surface brightness plofile of A\,2390 (B\"ohringer et al. 1998) fitted by our model
 (solid line) and the isothermal $\beta$ model (dushed line).}
 \label{fig:four}
\end{figure}

\end{document}